\documentclass{appolb}
\usepackage{graphicx}
\usepackage[utf8]{inputenc}
\usepackage{cite}

\begin{document}
\title{Mirror Neutron Stars: How QCD can be used to study dark matter through gravitational waves%
\thanks{Presented at Quark Matter 2022}%
}
\author{Maur\'icio Hippert, Hung Tan, Jacquelyn Noronha-Hostler, Nicol\'as Yunes
\address{Illinois Center for Advanced Studies of the Universe, Department of Physics, University of Illinois at Urbana-Champaign, Urbana, IL 61801, USA}
\\[3mm]
{Jack Setford, David Curtin 
\address{Department of Physics, University of Toronto, Toronto, ON M5S 1A7, Canada}
}
}
\maketitle
\begin{abstract}
Given the lack of empirical evidence of weakly interacting dark matter, it is reasonable to look to other candidates such as a confining dark sector with a similar number of particles as the standard model. Twin Higgs mirror matter is one such model that is a twin of the standard model with particles masses 3--6 times heavier than the standard model that solves the hierarchy problem. This generically predicts mirror neutron stars, degenerate objects made entirely of mirror nuclear matter. We find their structure using a realistic equation of state from crust (nuclei) to core (relativistic mean-field model) and scale the particle masses using lattice QCD results. We find that mirror neutron stars have unique signatures that are detectable via gravitational waves and binary pulsars, that provides an intriguing possibility for probing dark matter.
\end{abstract}
  
\section{Introduction}

Dark matter constitutes one of the great mysteries of present-day physics. 
While its existence seems to be firmly established by  astronomical observations and cosmologic data, dark matter continues to elude direct detection in particle-physics experiments. 
As traditional dark-matter candidates
fail to be detected and constraints from collider experiments 
become increasingly stringent, it is crucial that other possibilities for dark matter be considered that could produce entirely different experimental signatures. 
One such possibility is that dark matter could be \emph{complex} in nature, that is, composed from more than a single particle species, potentially interacting via their own force mediators. This idea of \emph{dark complexity} is not unnatural given the abundance of dark matter and the complex ``particle zoo'' of standard-model interactions. 
Moreover, complex dark matter candidates can also present solutions for pressing issues in the Standard Model (SM) of particle physics. 
This is the case of \emph{mirror matter}, comprised by exotic particles related to standard-model ones via a discrete symmetry and predicted by the \emph{Twin Higgs} mechanism, 
proposed as a solution to the \emph{Hierarchy Problem} of particle physics \cite{Chacko:2005pe}.

A generic and yet fascinating prediction of mirror matter, and of dark complexity in general, is the possibility of degenerate stars made entirely of dark matter \cite{Hippert:2021fch}. While these stars would have very faint or nonexistent visible signatures, collisions between sufficiently compact, massive objects might be discovered via the gravitational-wave radiation.  
In particular, mirror sectors featuring modified versions 
of QCD should lead to very dense dark objects akin to neutron stars, but made of mirror baryons. 
The discovery of a \emph{mirror neutron star} would revolutionize our understanding of dark matter, potentially revealing its very nature. 

Predictions for mirror neutron stars demand that knowledge from nuclear and neutron-star physics be extrapolated to the mirror sector. 
The minimal \emph{Mirror Twin Higgs} model predicts a copy of QCD with pion masses $2-4$ times larger than in the SM, lying in a range that has been extensively explored in past \emph{Lattice QCD} calculations \cite{Hanhart:2008mx,McNeile:2009mx,Walker-Loud:2008rui,LHPC:2010jcs,Syritsyn:2009mx,Albaladejo:2012te,Horsley:2013ayv}. 
This coincidence creates an exciting opportunity to employ knowledge from nuclear physics to study mirror neutron stars, with guidance from first-principles calculations \cite{Hippert:2021fch}. 

\section{Standard-Model Neutron Stars}

Dark-matter stars have been previously considered in the literature. 
However, for the first time \cite{Hippert:2021fch}, we build a complete description of standard-model neutron stars, capable of reproducing the latest observational constraints, before extrapolating this model to the mirror sector. 




The stellar crust is modelled as a lattice of nuclei embedded in an electron gas, according to the Baym-Pethick-Sutherland model \cite{Baym:1971pw}. The dominant nuclear species for each value of the pressure is determined by finding the most energetically favored combination of atomic number $Z$ and mass number $A$. The total energy per unit volume is the combination of contributions from the electron gas, the lattice interaction energy and the nuclear mass, with the nuclear binding energy $b_N(Z,A)$ given by a semi-empirical formula. 
As the density increases, heavier and more neutron-rich nuclei are found, which would not be stable outside an extremely dense environment. At the so-called neutron drip line, a neutron liquid starts forming around the nuclei and the model ceases to be valid.  




In the stellar core, we adopt a relativistic mean-field model. 
Above nuclear saturation density, our model for neutron-star matter consists of nucleons interacting via a scalar $\sigma$, a vector $\omega^\mu$ and a vector-isovector $\vec\rho^\mu$ mesons, plus interaction terms between mesons themselves \cite{Hippert:2021fch}. 
The couplings of this model are tuned by imposing that properties of nuclear matter at saturation be reproduced, and constraints from neutron-star observations be satisfied \cite{LIGOScientific:2017vwq,Miller:2019cac}. 
Local charge neutrality and chemical equilibrium under weak interactions are enforced by complementing the model with a gas of degenerate electrons and muons. 


 Between the neutron drip line and nuclear saturation density, we bypass complications such as the description of pasta phases by employing an interpolating function.  
The equations of state for the stellar crust and core are smoothly joined, so that the equation of state is continuous and the pressure and energy density increase monotonically with density \cite{Hippert:2021fch}. 


In the following, we show how this model for SM neutron-stars can be extended to describe mirror neutron-star matter.

\section{Mirror Twin Higgs Matter}

An interesting realization of mirror matter is provided by the {Twin Higgs} mechanism, proposed as a solution to the {Hierarchy Problem} regarding the mass of the Higgs boson \cite{Chacko:2005pe}. 
That is, as an explanation for why the Higgs mass is so small compared to the Plack scale, when it should receive very large quantum corrections from the ultraviolet. 
The {Twin Higgs} addresses this issue by invoking a hidden sector related to the SM by a discrete symmetry, which protects the Higgs mass from these large contributions. 

In the minimal \emph{Mirror Twin Higgs} model, hidden-sector particles and interactions are exact copies of the SM, but the mirror Higgs vacuum expectation value $f$ is larger than its SM counterpart $v$. 
As a result, all the new fundamental particles in the Mirror Twin Higgs model turn out to be heavier than their SM versions by the same factor $f/v$. 
For the model to evade current collider constraints and yet provide a satisfactory solution to the Hierarchy Problem,  the ratio between the Higgs and Mirror Higgs expectation values must lie in the range $f/v\sim 3-10$ \cite{Burdman:2014zta}. 

Thus, the minimal Mirror Twin Higgs provides us with a complete and well constrained model, from which predictions on mirror neutron stars can be made.

\section{Mirror Neutron Stars}

Results for mirror neutron stars demand that the parameters in our model are rescaled to account for the heavier current quark masses in the mirror sector. 
Because of the dynamical scale and chiral symmetry breaking of QCD, translating the change in current quark masses to changes in hadronic masses and couplings is not straightforward. 
However, the scaling of the mirror QCD scale $\Lambda_{\rm{QCD}}'$ with $f/v$ has been previously obtained from a fit to renormalization-group calculations  \cite{Burdman:2014zta}. They also allow us to calculate the scaling of the mirror pion mass $m_\pi'$:
  \begin{equation}\label{eq:scaling}
    \frac{\Lambda_{\rm{QCD}}'}{\Lambda_{\rm{QCD}}} \approx 0.68 + 0.41 \log\left(1.32 + \frac{f}{v}\right),\qquad
    m_\pi'\propto \sqrt{\Lambda_{\rm{QCD}}'\,m_q'}\,.
  \end{equation}
  

Parameters of the model for the stellar core can be extracted from lattice QCD calculations at heavier pion masses and chiral perturbation theory data \cite{Hanhart:2008mx,McNeile:2009mx,Walker-Loud:2008rui,LHPC:2010jcs,Syritsyn:2009mx,Albaladejo:2012te,Horsley:2013ayv}. 
When data is available, the scaling of each parameter with $(m_\pi'/\Lambda_{\rm{QCD}}')^2$ is fit to a quadratic function:
\begin{equation}
  \frac{m_i'}{\Lambda_{\rm{QCD}}'}
\approx b_0 + b_1 \left( \frac{m_\pi'^2}{\Lambda_{\rm{QCD}}'^2} \right) + b_2 \left( \frac{m_\pi'^2}{\Lambda_{\rm{QCD}}'^2} \right)^2 
    \end{equation}
Relations from low-energy nuclear physics are also employed to relate parameters to one another.

For the stellar crust, nuclear binding energies need to be adjusted according to the mirror QCD scale. We rescale them proportionally to $\Lambda_{\rm{QCD}}'$.     
Lepton masses, of course, scale in direct proportion to the mirror Higgs vacuum expectation value $m_\ell\propto f/v$. 

\section{Results}

\begin{figure}[htb]
\centerline{%
\includegraphics[width=8.5cm]{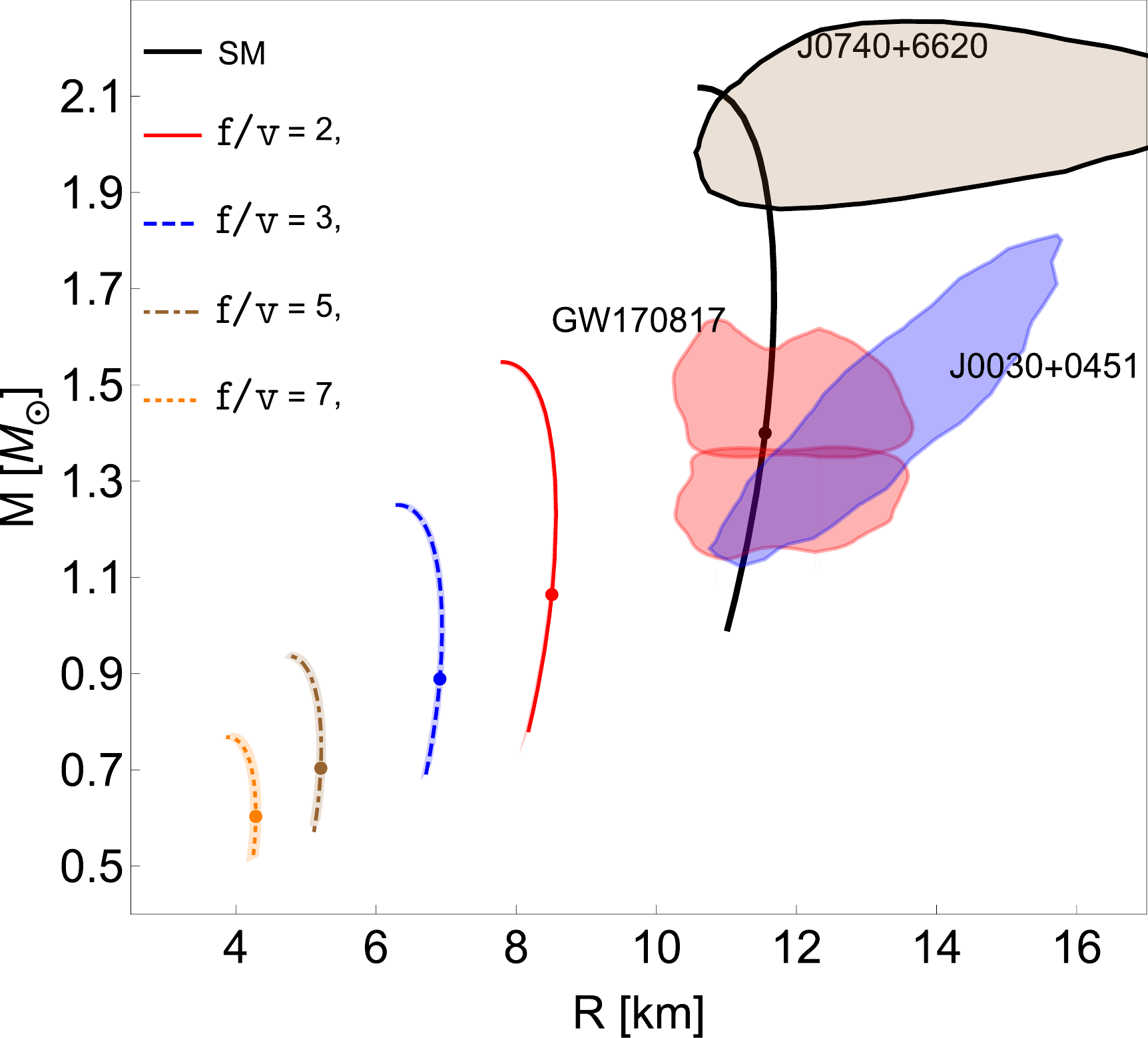}}
\caption{Mass-radius relations for standard-model (black curve) and mirror neutron stars (colored curves), compared to observational constraints \cite{LIGOScientific:2017vwq,Miller:2019cac,Miller:2021qha,Riley:2021pdl}.}
\label{Fig:MR}
\end{figure}

Our result for the mass-radius relation of standard-model neutron stars is shown as the black curve in Fig.~\ref{Fig:MR}, where they are compared to observational constraints. Parameters of our model were adjusted to reproduce constraints from gravitational-wave event GW170817 and the NICER observation of pulsar J0030+0451 \cite{LIGOScientific:2017vwq,Miller:2019cac}, but the agreement with NICER data for pulsar J0740+6620 is obtained as a \emph{prediction} \cite{Miller:2021qha,Riley:2021pdl}. 

Our predictions for the mass-radius relation of mirror neutron stars are shown, for different values of $f/v$, as the colored curves in Fig.~\ref{Fig:MR}. The phenomenologically relevant range $f/v = 3-7$ corresponds to the blue, brown and yellow curves. The thin bands around these curves represent uncertainties from fits to the lattice data. Stellar masses and radii can be shown to scale with ${m_B'}^{1.9}$, where $m_B'$ is the mirror baryon mass \cite{Hippert:2021fch}.

\section{Conclusions}

The mirror neutron stars predicted by the minimal Mirror Twin Higgs model inhabit a unique portion of the mass-radius diagram and are heavy enough to be discovered via binary merger events by advanced LIGO  \cite{Hippert:2021fch}. 
Together with their masses, the modest tidal deformabilities resulting from their small size would be enough to distinguish them from neutron stars, strange quark stars and black holes, by gravitational-wave imprints alone \cite{Hippert:2021fch}. Combined with the absence of electromagnetic signatures in mirror neutron-star collisions, these features amount to a promising discovery potential.

\section*{Acknowledgements}

This work was supported in part by the National Science Foundation (NSF) within the framework of the MUSES collaboration, under grant number OAC-2103680.
H.T. and N.Y.~acknowledge financial support through NSF grants  No.~PHY-1759615, PHY-1949838 and NASA ATP Grant No.~17-ATP17-0225,  No.~NNX16AB98G and No.~80NSSC17M0041. 
The research of J.S. and D.C. was supported in part by a Discovery Grant from the
Natural Sciences and Engineering Research Council of Canada, and by the Canada Research Chair
program. J.S. also acknowledges support from the University of Toronto Faculty of Arts and Science postdoctoral fellowship. 
J.N.H. acknowledges financial support by the US-DOE Nuclear Science Grant No. DESC0020633.
%


\end{document}